# Global room-temperature superconductivity in graphite

*Yakov Kopelevich*, José Torres, Robson da Silva, Felipe Oliveira, Maria Cristina Diamantini, Carlo Trugenberger, and Valerii Vinokur**

Y. V. Kopelevich*, J. H. S. Torres, R. R. da Silva, F. S. Oliveira
Universidade Estadual de Campinas-UNICAMP, Instituto de Física "Gleb Wataghin", R. Sergio Buarque de Holanda 777, 13083-859 Campinas, Brazil
kopel@ifi.unicamp.br, jhspahntorres@gmail.com, rorisi@ifi.unicamp.br, fso@unicamp.br

M. C. Diamantini
NiPS Laboratory, INFN and Dipartimento di Fisica e Geologia, University of Perugia, via A. Pascoli, I-06100 Perugia, Italy
cristina.diamantini@pg.infn.it

C. A. Trugenberger
SwissScientific Technologies SA, rue du Rhone 59, CH-1204 Geneva, Switzerland
ca.trugenberger@bluewin.ch

V. M. Vinokur*
Terra Quantum AG, Kornhausstrasse 25, CH-9000 St. Gallen, Switzerland
vv@terraquantum.swiss

**Abstract**

Room temperature superconductivity under normal conditions has been a major challenge of physics and material science since its very discovery. Here we report the global room-temperature superconductivity observed in cleaved highly oriented pyrolytic graphite carrying dense arrays of nearly parallel surface line defects. The multiterminal measurements performed at the ambient pressure in the temperature interval 4.5 K $\leq T \leq$ 300 K and at magnetic fields $0 \leq B \leq 9$ T applied perpendicular to the basal graphitic planes reveal that the superconducting critical current $I_c(T, B)$ is governed by the normal state resistance $R_N(T, B)$ so that $I_c(T, B)$ is proportional to $1/R_N(T, B)$. Magnetization $M(T, B)$ measurements of superconducting screening and hysteresis loops together with the critical current oscillations with temperature that are characteristic for superconductor-ferromagnet-superconductor Josephson chains, provide strong support for occurrence of superconductivity at $T > 300$ K. We develop a theory of global superconductivity emerging in the array of linear structural defects which well describes the experimental findings and demonstrate that global superconductivity arises as a global phase coherence of superconducting granules in linear defects promoted by the stabilizing effect of underlying Bernal graphite via tunneling coupling to the 3D material.

## 1. Introduction

The discovery of superconductivity (SC) in mercury at 4.2 K[1] triggered a dream of superconductivity at room temperature, realizing which has now become one of the major tasks of physics and material science. An extensive search for room temperature superconductivity (RTSC) is motivated both by fundamental appeal and by the exclusive platform that RTSC offers for broad technological applications. While several systems have demonstrated close to room temperature SC under high pressures,[2, 3] its observation under ambient conditions still remains a challenge. The discovery of high-temperature superconductivity (HTSC) in the Ba-La-Cu-O cuprates with $T_c \approx 30$ K[4] and Y-Ba-Cu-O with $T_c$ being as high as 93 K[5] marked a breakthrough in the RTSC search and brought in a hope for its fast coming. So far, the mercury-based cuprate $HgBa_2Ca_2Cu_3O_9$ showed the highest $T_c = 135$ K under the ambient pressure.[6]

Graphite is yet another promising material taking part in a race for the RTSC. Decades ago, Antonowicz[7] measured the Josephson-type oscillations and Shapiro-like steps in current-voltage, *I*-*V*, characteristics at $T = 300$ K in Al-AC-Al sandwiches, where the AC stands for the amorphous carbon. Various experimental groups have also reported localized superconductivity in graphite at temperatures as high as 300 K.[8, 9] Because the AC consists of curved graphene and/or of fullerene-like fragments, one can justly assume that similar structural defects in graphite may be responsible for the occurrence of high-temperature localized superconducting regions. However, so far, all the efforts to achieve a global superconductivity at elevated temperatures in graphite failed.

In the present work, we report the first unambiguous experimental evidence for the global zero-resistance state, RTSC, in the scotch-tape cleaved highly oriented pyrolytic graphite (HOPG) that possesses dense arrays of nearly parallel line defects (LD), the wrinkles.

## 2. Experimental setting

In our experiments, we use the scotch-taped cleaved pyrolytic graphite carrying the wrinkles that resulted from this cleaving to which we also refer as to line defects (LD). The surface carries the bundles of the narrow-separated wrinkles, see Fig. 1a, with the bundles separated from each other by the distance of $d = 0.2$ mm, as shown in **Figure 1a**; each bundle is drawn as a line in the low **Figure 1a** panel.

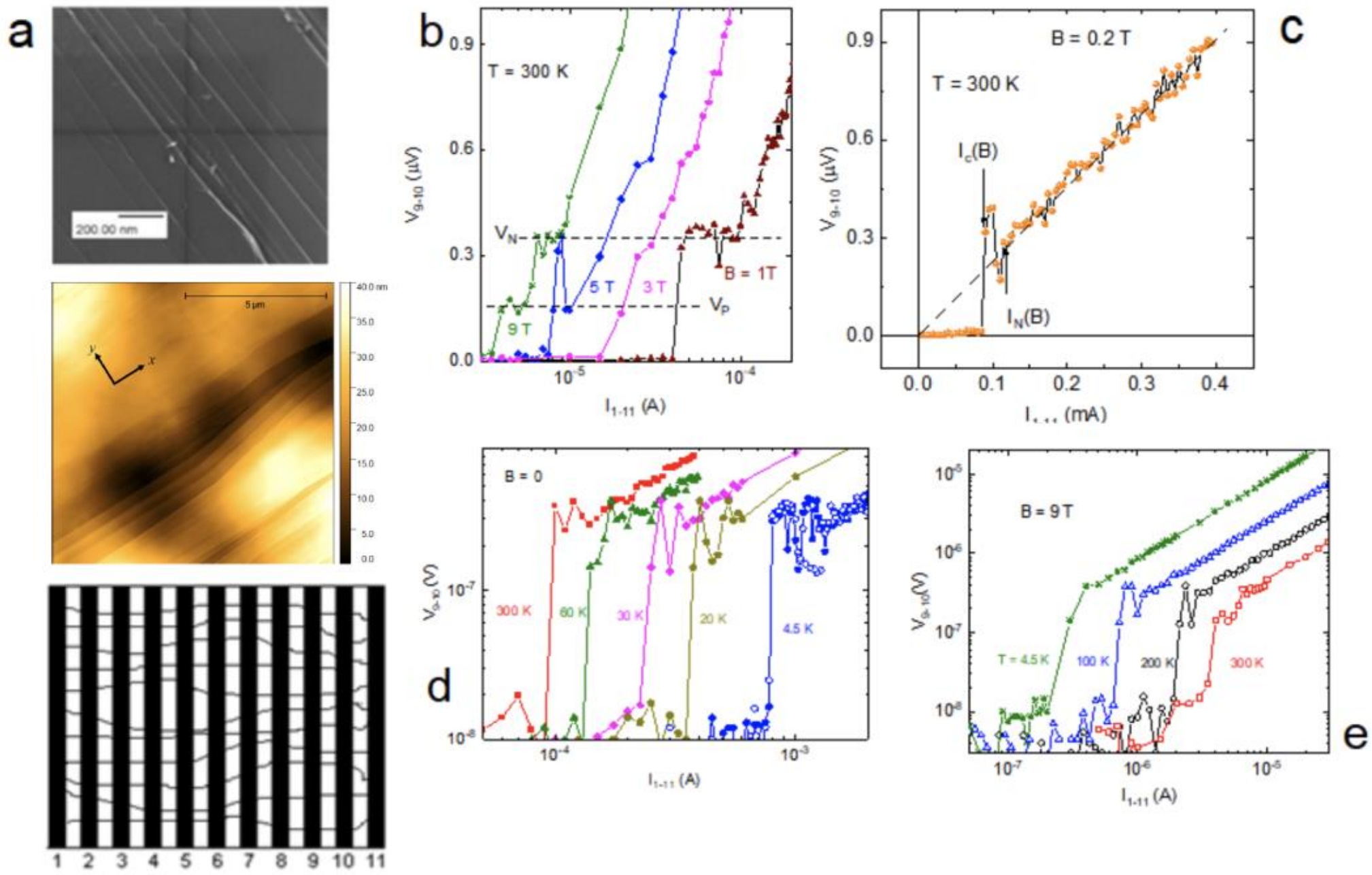

**Figure 1. a:** Upper panel: Scanning electron microscope (SEM) image of the HOPG sample demonstrating large scale nearly parallel line defects. Middle panel: Atomic Force Microscope (AFM) image, demonstrating the ridge-type topography of the sample surface. Lower panel: A sketch of the device. The numbers correspond to eleven silver epoxy electrodes, shown as black bars, separated by the distance $d = 0.2$ mm and disposed on the cleaved surface of the HOPG sample. Thin curved lines sketch the line defects. **b:** Semilogarithmic plot of *I*-*V* characteristics measured at $T = 300$ K for selected applied magnetic fields; $V_N = 0.36\ \mu V \approx 2V_P$. **c:** The *I*-*V* curve in a linear scale obtained for $B = 0.2$ T; the dashed line corresponds to the Ohm's law. The $I_c(T, B)$ and $I_N(T, B)$ dependences present currents corresponding to transitions to superconducting and normal states, respectively. **d:** Selected *I*-*V* characteristics, measured at $B = 0$ and $4.5\ K \leq T \leq 300$ K. The curves demonstrate decrease of the critical current $I_c(T)$ with the temperature increasing. Measurements at $T = 4.5$ K were performed for both increasing (•) and decreasing (o) current, testifying the reversible character of *I*-*V* curves. **e:** The I-V characteristics, measured for $B = 9$ T and $4.5\ K \leq T \leq 300$ K demonstrate the increase of the critical current $I_c(T)$ with the increasing temperature.

In experiments, we use the line-electrode geometry to measure the in-plane resistance in both local and non-local configurations. Eleven silver epoxy electrodes with the contact resistance $R_c \approx 1\ \Omega$ oriented perpendicular to the wrinkles, are patterned on one of the main surfaces of the graphite sample with the separation distance $d = 0.2$ mm, as shown in **Figure 1a**. The spatially resolved micro-Raman measurements[10] performed on the cleaved HOPG-UC samples with the wrinkles revealed that depending on the laser spot location on the sample surface, one observes either the characteristic of the bulk graphite left-shoulder-2D Raman peak or the additional peak, characteristic for the single- or multi-layer graphene (MLG). This experimental fact indicates the existence of either independent or weakly coupled graphene layers in the wrinkles. Supporting such a conclusion, the Raman spectra measured for the MLG flakes with the LD[11] show that the interlayer coupling at a wrinkle is weaker than that in the flat regions.

The dc current is applied either between the current leads 1 and 11, $I_{1\text{-}11}$ or leads 1 and 4, $I_{1\text{-}4}$. In the first configuration, we measure the voltages $V_{23}\ldots V_{10\text{-}11}$ in the current-applied region, which we refer to as to local voltages. In the second case, the voltage drops were measured simultaneously in both the applied current part of the crystal $V_{23}$ and outside that region, $V_{56}\ldots V_{10\text{-}11}$, the latter we refer to as to the non-local voltages. Here we report the results obtained for both the local, $I_{1\text{-}11}$-$V_{9\text{-}10}$, measurement configuration and the results for the current applied between the leads 1 and 4, i.e., $I_{1\text{-}4}$, with $I_{1\text{-}4} \equiv I_0$=10 mA, and the voltages measured outside the region between the leads 1 and 4. Transport measurements are performed for the $B$||c-axis.

## 3. Results

### 3.1. Resistance measurements

Figure 1b,c presents the *I*-*V* characteristics measured at $T = 300$ K. The data demonstrate the zero-resistance state below the magnetic-field-dependent critical current $I_c(B)$, which is decreasing with the field. The obtained *I*-*V* curves demonstrate the characteristic features of low-dimensional superconductors. First, the excess voltage peaks seen just above the $I_c$(B) and before the Ohmic regime sets in at $I > I_N$, see Fig. 1c, are similar to those measured in one (1D)- or two (2D)-dimensional superconducting constrictions, and are attributed to the charge imbalance and/or presence of phase slip (PS) centers at superconductor (S) - normal metal (N) interfaces.[12] The onset of the Ohmic behavior in *I*-*V* characteristics corresponds to the suppression of the non-equilibrium superconducting regime or the transition to the normal state.

Figure 1b also demonstrates the appearance of voltage plateaus in the *I*-*V* curves. These plateaus develop at two voltage levels, viz., at the normal state voltage $V_N$ and at $V_P \approx V_N/2$. A similar plateau at $V_P \approx V_N/2$ has been reported for low-$T_c$ superconducting nanowires in a non-hysteretic out-of-equilibrium dissipative regime.[13]

The *I*-*V* characteristics measurements reveal a qualitatively different $I_c(T, B)$ behavior below and above the crossover field $B_x \approx 35 \pm 5$ mT, which separates $dI_c/dT < 0$ and $dI_c/dT > 0$ behaviors for $B < B_x$ and $B > B_x$, respectively. Figures 1d and 1e demonstrate this behavior for the *I-V* curves measured for $B = 0$, see Fig. 1d, and $B = 9$ T, see Fig. 1e, for some selected temperatures.

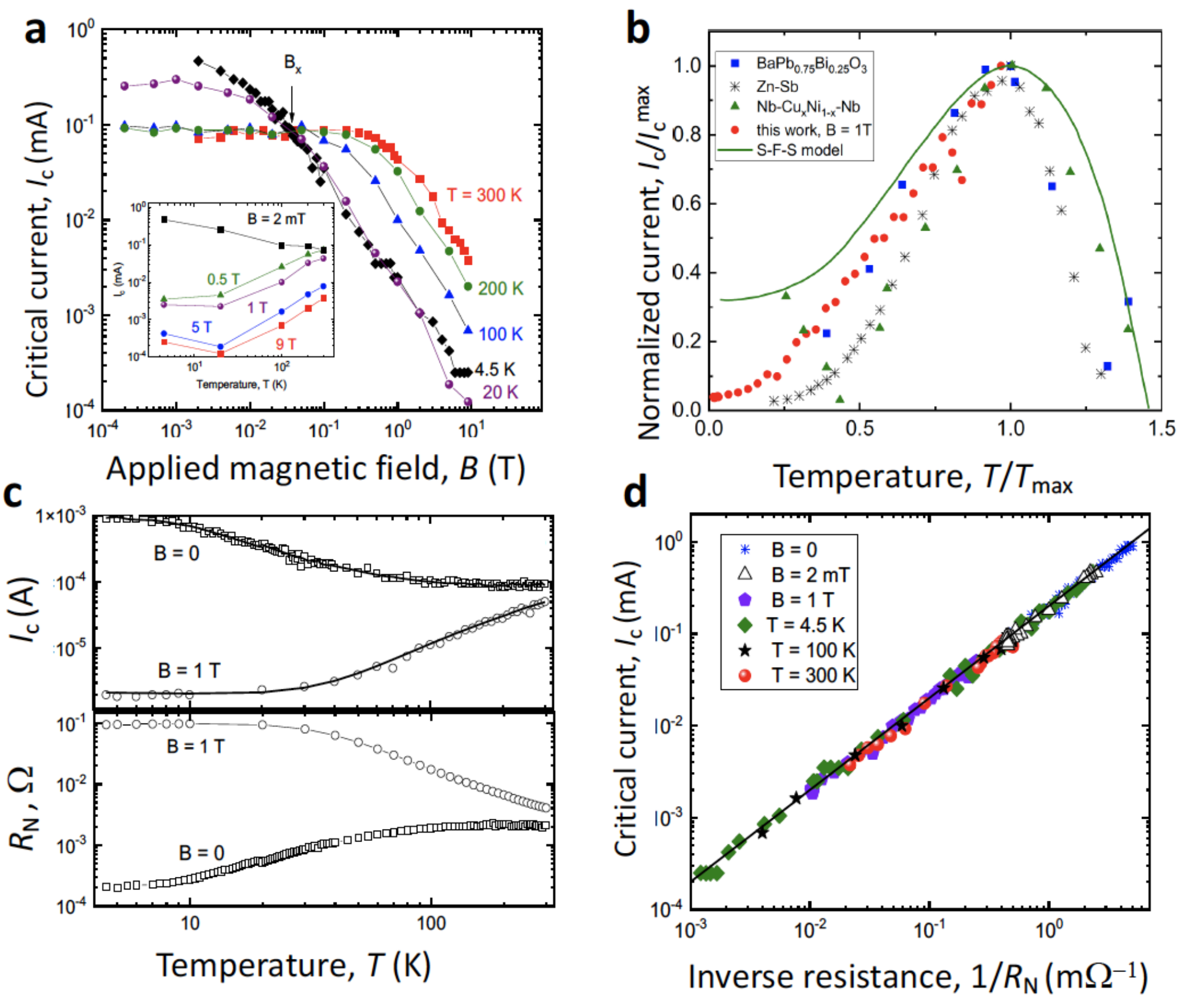

**Figure 2. The temperature and magnetic field dependencies of the critical current. a:** Magnetic field dependencies of the critical current $I_c(B, T)$ extracted from the *I-V* isotherms. The crossover field $B_x \approx 35$ mT separates $dI_c/dT < 0$ ($B < B_x$) and $dI_c/dT > 0$ ($B > B_x$) behaviors. The inset illustrates $I_c(T)$ obtained for some selected magnetic fields from the main figure. **b:** Reduced $I_c(T)$ vs. reduced $T$ for $BaPb_{0.75}Bi_{0.25}O_3$ [15], Zn-Sb [16], Nb-$Cu_xNi_{1-x}$-Nb SC-FM-SC Josephson junctions,[18] and graphite (this work). The critical current $I_c(T)$ reaches the maximum value $I_c^{max}$ at $T = T_{max}$. The model calculation of $I_c(T)$ for the SC-FM-SC) Josephson junction is represented by the solid line.[19] Our data for $I_c(T$, B = 1 T) are the same as in panel a taking $T_{max} = 300$ K**. c, upper panel:** $I_c(T, B)$ obtained from *I-V* characteristics for 4.5 K ≤ $T$ ≤ 300 K at $B = 0$ and $B = 1$ T. Solid lines are obtained from the equation $I_c = V_c/R_N(T, B)$, $V_c = 0.2$ μV, and $R_N(T, B)$ is the normal state resistance. **c, lower panel:** $R_N(T, B)$ for $B$=0 and $B = 1$ T for $I > I_c(T, B)$ in the Ohmic regime. d: $I_c(T, B)$ vs. $R_N(T, B)$ obtained from *I-V* characteristics measured at a fixed temperature, $T = 4.5$, 100, and 300 K or fixed magnetic field, $B = 0$, 2 mT, and 1 T. The solid line corresponds to the equation $I_c = V_c/R_N$ with $V_c = 0.2$ μV.

**Figure 2a** summarizes the results of the $I_c(T, B)$ measurements. It reveals the existence of the crossover field $B_x \approx 35$ mT that separates the $dI_c/dT < 0$ behavior typical for most superconductors at $B < B_x$, and anomalous behavior with $dI_c/dT > 0$ at $B > B_x$. To better visualize the anomalous behavior, we plot $I_c(T, B)$ for a few selected fields in the inset in Figure 2a. Note that the $dI_c/dT > 0$ behavior is well known for type-II superconductors and is mostly observed either in the vicinity of the upper critical field $B_{c2}(T)$ or just below the Abrikosov vortex lattice melting phase transition.[14] That is, it is usually seen only within a narrow interval of magnetic field and temperature. In our case, e.g., $I_c(T, B = 9$ T) rises about ten times as the temperature increases from 4.5 K to 300 K, see the inset in Figure 2a.

At the same time, a broad maximum in $I_c(T)$ and $dI_c/dT > 0$ for $T \ll T_c$ was reported for various granular and inhomogeneous superconductors, such as, for instance, $BaPb_{0.75}Bi_{0.25}O_3$ granular superconductor,[15] Zn-Sb inhomogeneous alloys,[16] Sn-$SnO_x$-Sn tunnel Josephson junctions (JJ),[17] and for superconductor-ferromagnet-superconductor (SC-FM-SC) JJ.[18] In Figure 2b we plot the normalized critical current $I_c(T)/I_c(T=T_{max})$ vs. $T/T_{max}$ for $BaPb_{0.75}Bi_{0.25}O_3$,[15] Zn-Sb,[16] and Nb-$Cu_xNi_{1-x}$-Nb JJ[16] together with our data for $I_c(T, B)$ obtained for $B = 1$ T $\gg B_x$, where $T_{max}$ corresponds to the maximal value of $I_c$. As seen from this panel, $I_c(T)$ dependencies are quite similar for all these systems. Assuming that $I_c(T, B = 1$ T) in our case reaches maximum at $T_{max} \geq 300$ K, which is the highest measuring temperature, one gets surprisingly good matching between our and the literature data suggesting that the Josephson junction array-like medium is indeed behind of the $I_c(T)$ behavior in all these materials. We have also found that the $I_c(T, B)$ behavior can be fully described by using the temperature and magnetic field dependences of the normal state resistance $R_N(T, B)$ as expected for the JJ critical current.

Various mechanisms to explain the non-monotonic $I_c(T)$ behavior have been suggested, such as (i) the reduction of the charge carrier density upon lowering temperature in the superconductor-semiconductor-superconductor JJ;[15] (ii) temperature dependence of the sub-gap quasi-particle conductance;[17] (iii) a non-uniform current distribution;[16] and (iv) crossover between the $\pi$- and zero-phase superconducting states in SC-FM-SC JJ.[18,19] The last scenario is particularly appealing in our case of a few-layer graphene LD on graphite, where the coexistence of superconductivity and magnetism and their interplay is well confirmed experimentally.[8,20-22] The upper panel of Figure 2c presents $I_c(T)$ for $B = 0$ and $B = 1$ T for 4.5 K $\leq T \leq$ 300 K. The experimental data can be nicely fitted by solid lines obtained from the equation $I_c(T, B) = V_c/R_N(T,B)$, where $V_c = 0.2$ μV. The resistance, $R_N(T, B)$, temperature dependences measured for $B = 0$ and $B = 1$ T, are shown in the lower panel of Fig. 2c. One sees that the crossover from the conventional, $dI_c/dT < 0$ for B = 0, to anomalous $dI_c/dT > 0$ for $B = 1$ T behavior is governed by the field-induced transformation from the metallic-like $dR_N/dT > 0$, at $B < B_x$, to the insulator-like $dR_N/dT < 0$, at $B > B_x$, resistance behavior.

Figure 2d illustrates the universality of the equation $I_cR_N = V_c$. One sees that $I_c$ vs. $1/R_N$ dependencies obtained for various $T$ and $B$ collapse on a single line $I_c = V_c/R_N$ with $V_c = 0.2\mu$V, spanning about four orders of magnitude in both $I_c(T, B)$ and $R_N(T, B)$ dependences. This remarkable invariance immediately suggests that $I_c$ and $R_N$ can be associated with the critical current and normal resistance of a Josephson junction, where the product $I_cR_N$ depends only on the properties of the materials involved but not on the geometry or dimensionality of a junction.[23] Microscopic derivation in the framework of the BCS theory gives well known result $I_c = (\pi\Delta/2eR_N)\cdot\tanh(\Delta/2k_BT)$, where $\Delta$ is the superconducting gap and $k_B$ is the Boltzmann constant.[24] Accordingly, $I_c(T{=}0) = \pi\Delta(0)/2eR_N$ and $I_c(T_c/2) \approx 0.9I_c(0)$, where $\Delta(0)$ is the magnitude of the superconducting gap at zero temperature, and $R_N$ is the JJ resistance just above $T_c$. However, taking the experimental values of $I_c = 9\cdot10^{-4}$ A and $R_N = 2\cdot10^{-4}$ $\Omega$ for $T = 4.5$ K and $B = 0$, one gets $\Delta(0) \approx 10^{-7}$ eV which is by many orders of magnitude much too small to account for the experimental results. This can be taken as an indication that either the pairing mechanism is by no means the BCS one or that some strong depression of the $I_cR_N$ product occurs, or both.

To better characterize the superconducting transition, we carry out measurements of the resistance as function of temperature, magnetic field, and applied current. Figure 3 presents the voltage $V(T)$ and the resistance $R(T) = V(T)/I$ plots for various measuring currents and $B = 0$. We see that the superconducting transition temperature $T_c(I)$ decreases as the current increases, and that for $I \geq 2$ mA no transition is seen down to $T = 4.5$ K. Figure 3 also demonstrates several salient features of the transition: (i) the two-step transition towards the superconducting state; (ii) voltage plateaus at $V_P = V_N/2$ within the current-dependent temperature interval $\Delta T(I)$; and (iii) the stochastic switching between the superconducting state, $V = 0$, the voltage plateau with $V = V_P$, and the normal state. Figure 3b provides the evidence for the Ohmic resistance $R(T)$ for $I \geq 2$ mA. The inset in Figure 3b presenting the $I_c(T)$ together with the data of Figure 2c for $I_c(T, B = 0)$ shows good agreement between these two independent measurements. Figure 4a shows temperature dependencies of the resistance $R(T, B)$ for various applied fields $0 \leq B \leq 40$ mT and the current $I = 0.5$ mA, which is slightly below $I_c(4.5\text{ K}, B = 0)$, together with $R(T, B)$ obtained for $I = 5$ mA $> I_c$. The curves $R(T, B)$ obtained for $B = 0$, 0.5, and 1 mT demonstrate sharp superconducting transitions, as well as the stochastic switching between the superconducting state, the intermediate state, having resistance $R \approx R_N/2$, and the normal state.

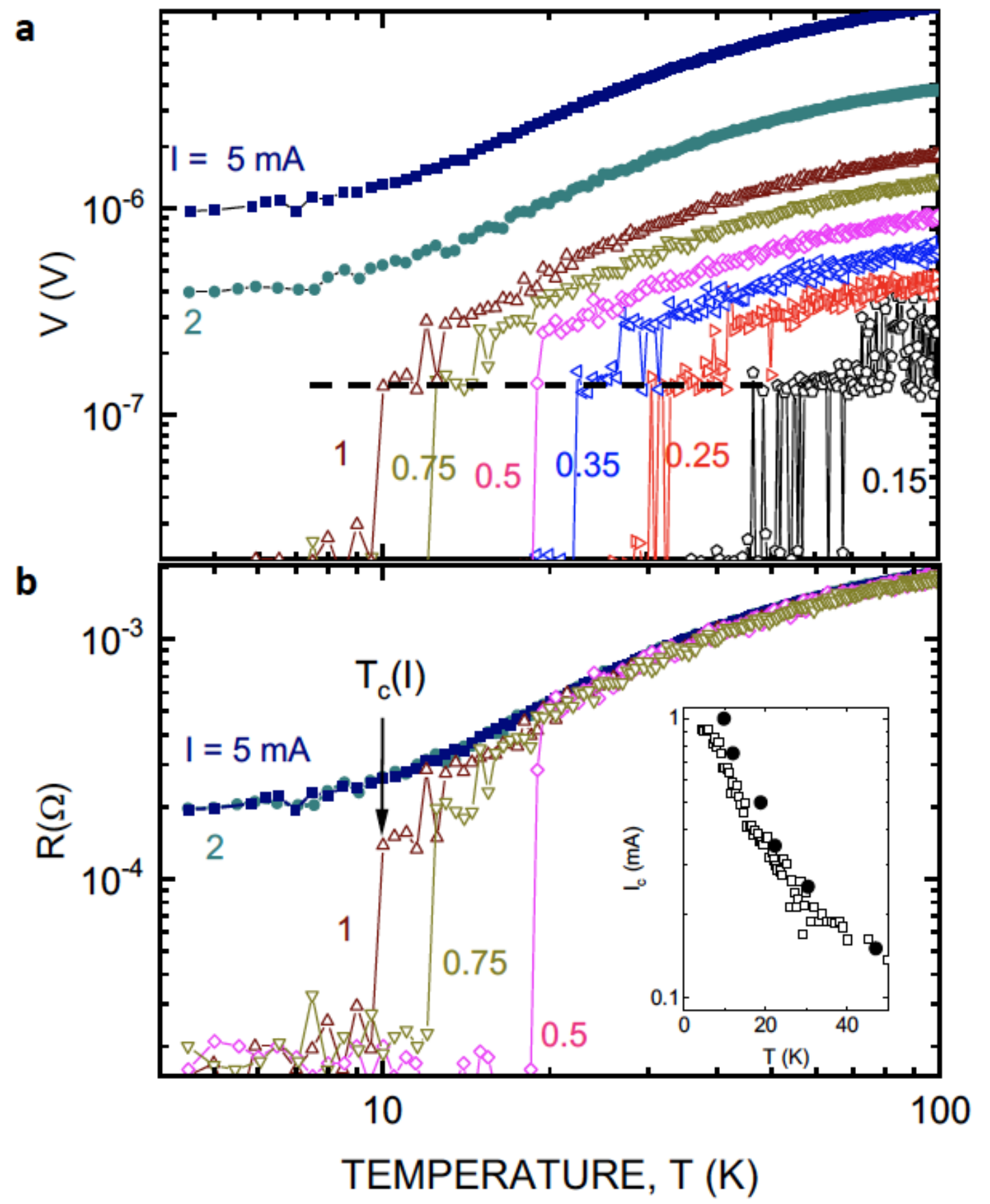

**Figure 3. Superconducting transition. a:** $V = V_{9\text{-}10}(T, B = 0)$ measured at various currents $I = I_{1\text{-}11}$ at $B = 0$ varying from $I$=5μA to $I$=0.15μA; the dashed line corresponds to the voltage level $V = 0.15\mu$V. **b:** $R(T) = V(T)/I$ obtained from (**a**) for these selected currents; the inset presents $I_c(T)$ (•) together with the data of Fig.2c (□ ) for $I_c(T,B = 0)$.

Figure 4a further illustrates that the resistance switching is not observable at temperatures below ≈ 18.5 K for $B = 0$, as well as at $T \leq 10$ K for $B = 0.5$ mT, and at $T \leq 8.5$ K for $B = 1$ mT, revealing the magnetic field effect on the stability of both the superconducting and the resistive states. For $B \geq 10$ mT, the sample is in the normal Ohmic state, as evident from the resistance measurements at $I = 5$ mA shown by solid lines. Further increase of $B$ results in the upturn,

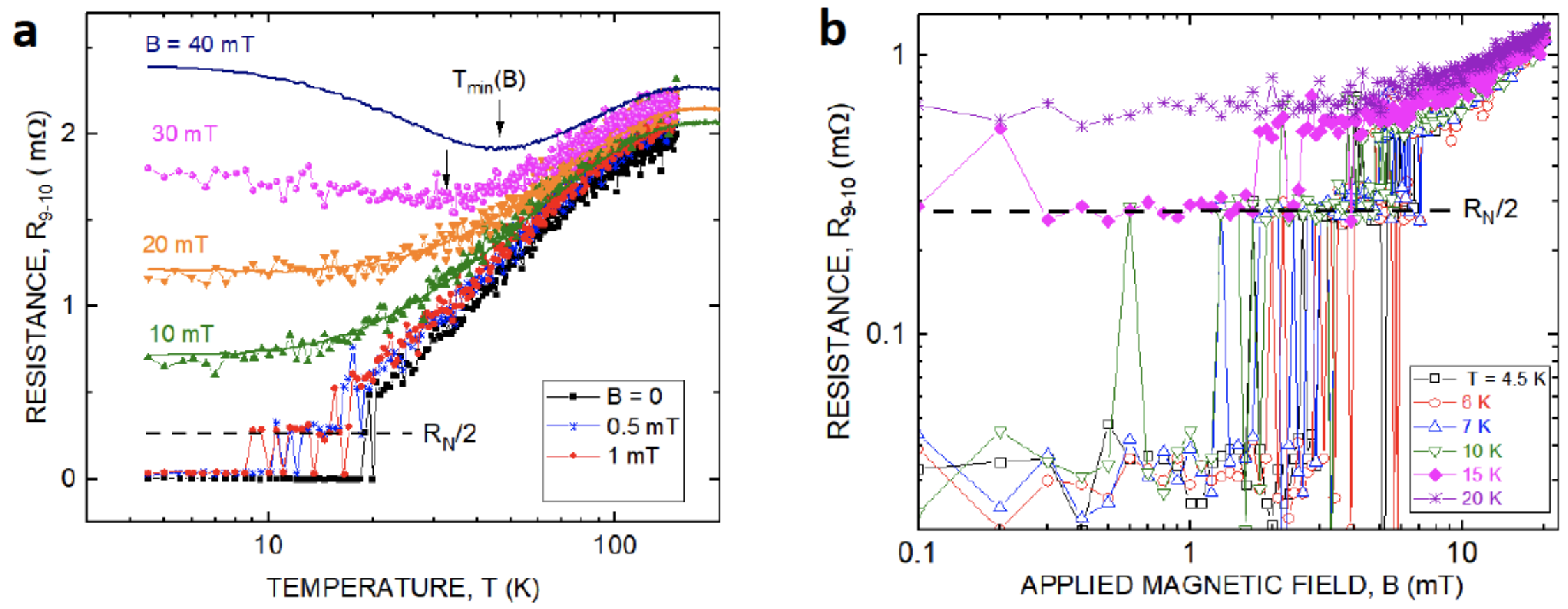

**Figure 4. Temperature and magnetic field dependences of the resistance. a:** Temperature dependences of the resistance $R \equiv R_{9\text{-}10}(T, B)$ measured with $I_{1\text{-}11} = 0.5$ mA (symbols) for various magnetic fields $0 \leq B \leq 40$ mT. Solid lines depict $R(T, B)$ measured for $I = 5$ mA $> I_c$. Arrows mark $T_{min}(B)$ corresponding to the minima in $R(T, B)$ curves. **b:** Magnetic field dependences of the resistance $R_{9\text{-}10}(B, T)$ measured with $I_{1\text{-}11} = 0.5$ mA for selected temperatures. Dashed lines in both graphs correspond to $R_N/2 \approx 0.28$ mΩ, the half of the normal state resistance.

$dR/dT < 0$, behavior for $T < T_{min}(B = 30\text{ mT}) = 37 \pm 1$ K, and $T < T_{min}(B = 40\text{ mT}) = 45 \pm 1$ K. The results shown in Figure 4a resemble the magnetic-field-driven superconductor-insulator transition (SIT) observed in two-dimensional (2D) Josephson junction arrays (JJA)[25] and in superconducting films[26] as well as in arrays of the 1D superconducting wires.[27] Figure 4b presents the magnetoresistance (MR) $R(B, T)$ obtained for the same measuring current $I = 0.5$ mA at several temperatures. Similar to the $R(T)$ dependences, the bistability regime in the magnetoresistance between zero- and the intermediate resistance state with $R \approx R_N/2 \approx 0.28 \pm 0.01$ mΩ becomes clear for $B$ slightly above 1 mT. For $T \geq 15$ K, only the dissipative intermediate superconducting state is visible, and at $T \geq 20$ K, the sample is in the normal state. Surprisingly, the measured resistance is independent of the applied magnetic field spanning almost two orders of magnitude, as can be clearly seen for $R(B, T = 15\text{ K})$.

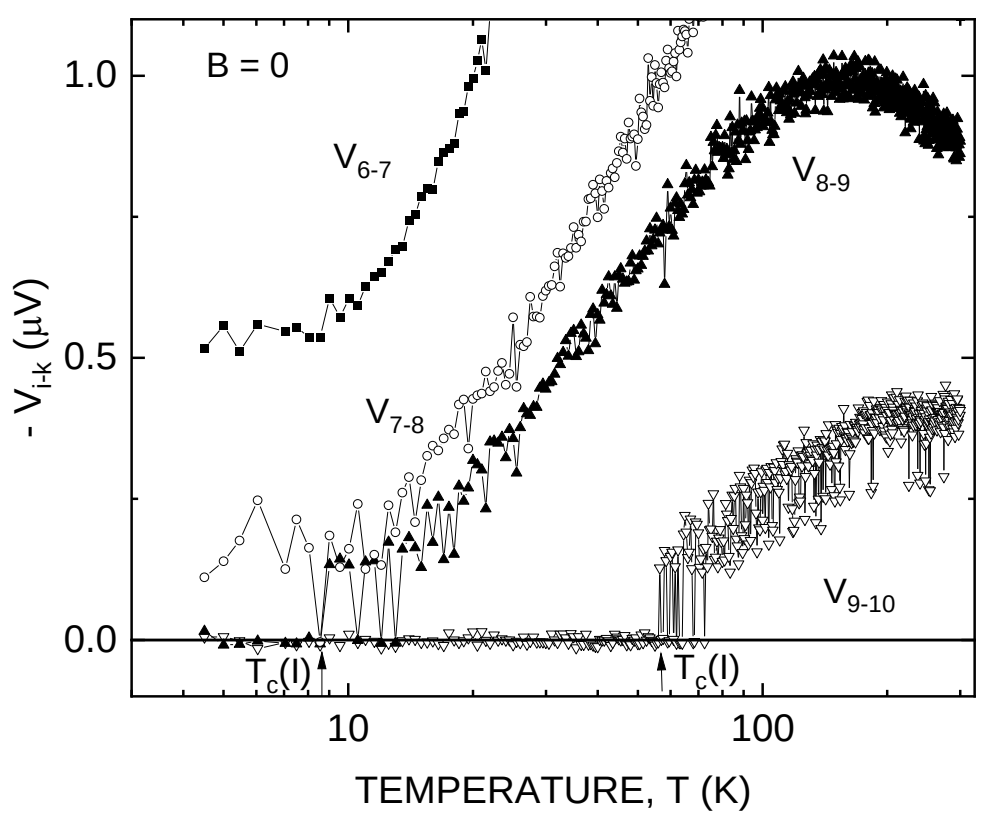

Figure 5. Temperature dependencies of the voltages $V_{i\text{-}k}$ measured in the "non-local" configuration with current applied between leads 1 and 4, $I_{1\text{-}4} = 10$ mA, and zero magnetic field, $B = 0$.

It is instructive to compare the data obtained for the voltage from the 9-10 electrode pair and the current sent across the whole sample from electrode 1 to electrode 11 with the data obtained from the arrangement when the current is sent through electrodes 1 and 4. Figure 5 illustrates these data. It presents the temperature dependencies of the voltages $V_{i\text{-}k}$ measured in the "non-local" configuration with the current applied between leads 1 and 4, $I_0 = I_{1\text{-}4} = 10$ mA. In this current in-and-out geometry, the "non-local" voltage is related to the current flowing outside of the electrodes 1-4 region, i.e., in fact, it has a local origin, see, for example, Refs. [28,29].

Outside the region between the electrodes 1-4, the current decays exponentially with the distance $x$ from the lead 4, $I(x) = I_0\exp(-x/x_0)$ with $x_0 = 0.5$ mm. This gives $I_{6\text{-}7} \approx 1.3$ mA, $I_{7\text{-}8} \approx 0.6$ mA, $I_{8\text{-}9} \approx 0.27$ mA, and $I_{9\text{-}10} \approx 0.12$ mA. Figure 5 and Fig.3a demonstrate a nearly perfect quantitative agreement between the superconducting transitions (temperature/current) obtained from the measurement of $V_{9\text{-}10}(T)$ in "local" arrangement with current $I_{1\text{-}11} = 0.15$ mA and "non-local," $I_{1\text{-}4}$, current configurations. The bistability regime followed by the zero-resistance state takes place below $T_l = 67$ K, Fig. 3(a), and $T_{nl} = 72$ K, seen in Fig.5, in "local" and "non-local" measuring schemes, respectively. The temperature intervals for the bistability also nearly coincide, as $T_{nl} \approx 16$ K and d$T_{l} \approx 20$ K.

At the same time, the superconducting transition temperature determined from the "non-local" voltage $V_{8\text{-}9}(T)$ with $I_{8\text{-}9} \approx 0.27$ mA is $T_{c}(x) \approx 6$ K, which is 5 times lower than $T_{c}$ obtained from the $V_{9\text{-}10}(T)$ local measurements for the same current value. Already, in the "nonlocal" $V_{7\text{-}8}(T)$ with $I_{7\text{-}8} = 0.6$ mA measurements, the superconducting transition was not observed down to our lowest measuring temperature $T = 4.5$ K. Once again, the superconducting transition temperature obtained from the local $V_{9\text{-}10}(T)$ measurements for the same current amplitude $I$ = 0.6 mA is $T_{c}(x) \approx 15$ K. These results indicate that the superconducting transition temperature, $T_{c}$, depends not only on the amplitude of the applied current but also on a spatial variation of the graphite properties that control the global superconductivity.

One of the central experimental observations of our work is the critical current dependence on the normal state resistance is $I_{c}(\mathrm{T, B}) \approx 1/R_{N}(\mathrm{T, B})$. To verify whether the results presented in Fig. 5 are consistent with such a correlation, we plot in Fig. 6 the temperature dependencies of the normal state resistance $R_{ik}(T) = V_{ik}(T)/I_{1\text{-}11}$, obtained from the measurements in the local configuration, with the current flowing between electrodes 1 and 11, with $I_{1\text{-}11} = 5$ mA. Figure 6 demonstrates that (i) the residual, R(T= 4.5 K) ), normal state resistances R$R_{6\text{-}7}$ = 1.64 mΩ, R$R_{7\text{-}8}$ = 0.66 mΩ, R$R_{8\text{-}9}$ = 1 mΩ, R$R_{9\text{-}10}$ = 0.33 mΩ, and (ii) that the residual resistance ratios RRR = $R(T$=170 K)/$R(T$=4.5K) are: RR$R_{6\text{-}7}$ = 1.36, RR$R_{7\text{-}8}$ = 1.84, RR$R_{8\text{-}9}$ = 1.6, RR$R_{9\text{-}10}$ = 7.0. Because $R_{9\text{-}10}(T)$ possesses the lowest residual resistance and is several times bigger than RRR, we expect higher $T_{c}(I)$ for the 9-10 sample part, which appears to be in excellent agreement with the experiment.

Indeed, $T_{c}(I = 0.27$ mA$) \approx 8.6$ K, see Fig. 5, obtained from the $V_{8\text{-}9}(T)$ dependence, is lower by a factor of approximately 3 as compared to $T_{c} \approx 28$ K obtained for the same current amplitude from the local $V_{9\text{-}10}(T)$ measurements shown in the Fig. 3. The obtained difference in $T_{c}$ can be accounted for by the RR ratio R$R_{89}$/R$R_{9\text{-}10} \approx 3$. It is also important to compare the results obtained for $V_{8\text{-}9}(T)$ and $V_{7\text{-}8}(T)$: while R$R_{7\text{-}8}$/R$R_{8\text{-}9} \approx 0.66$, no superconducting transition is seen in $V_{7\text{-}8}(T)$ dependences down to $T = 4.5$ K because of the higher current, $I_{7\text{-}8} = 0.6$ mA flowing between electrodes 7 and 8, such that $I_{8\text{-}9}/I_{7\text{-}8} \approx 0.45$. This demonstrates not only the quantitative agreement between the results obtained for different electrode pairs but also provides additional support to our conclusions about the origin of the global superconducting phase coherence in graphite with linear defects.

It is well known that the basal-plane resistance $R(T)$ in the HOPG samples is sensitive to the mosaicity, i.e., to the mutual misorientation of the small, l x w x t = 10 μm x 10 μm x 100 nm size blocks with the graphite crystallographic c-axis parallel to the block thickness. The mosaicity can be measured by means of the X-ray rocking curve and quantified by the full width at the half maximum (FWHM). In our HOPG sample, the FWHM ≈ 0.3° and it places it in the group of the best quality HOPG samples. However, the spatial variation of the FWHM on a mm scale is practically unavoidable. This should be the main source for the $R(T)$ spatial variation. The better metallicity measured for $V_{9\text{-}10}$, see Fig. 6, means the better graphite crystallite orientation. That is why we chose to carry out our measurements mainly for this 9-10 pair of electrodes, where the supercurrent mainly flows.

The density of the linear defects on the plain is about $10^4$/mm, which gives the effective linear defects cross-section $S \approx 10^{-15}$–$10^{-13}$ m$^2$, and we find from the value $I_c(T=0, B=0) = 1$ mA, as determined from Fig. 3, that the superconducting critical current density $j_c(T=0, B=0) \approx I_c/S \approx 10^6 - 10^8$ A/cm$^2$, which is characteristic to many standard superconductors[23].

Next, we briefly discuss the role of the line defects in the formation of the RTSC. The experimental evidences accumulated during the past two decades indicate that both SC and the competing ferromagnetic FM orders in graphite observed even above the room temperature are related to structural defects.[8,30,31] Linear defects comprise rich morphology of the disordered graphitic structures such as corrugations and wrinkles,[11] steps of the height $h_0 \leq h \leq 5h_0$, where $h_0 = 3.35$Å is the distance between the neighboring graphene planes in graphite,[30] folded multi-layer graphite/graphene, and ridges and graphitic strands,[32] Our results of atomic force microscopy (AFM) measurements, presented in Fig.1(a), demonstrate that ridges are characteristic topographic features of cleaved HOPG samples. This has been observed quite a long time ago; see, for example, Refs. [30] and [33], and references therein. The results reported in[33] revealed the strain-induced pseudo-magnetic field of 230 T and related to it Landau-level-like quantization at 77 K along a cleavage-induced ridge. This experimental fact strongly corroborates our theoretical results presented below.

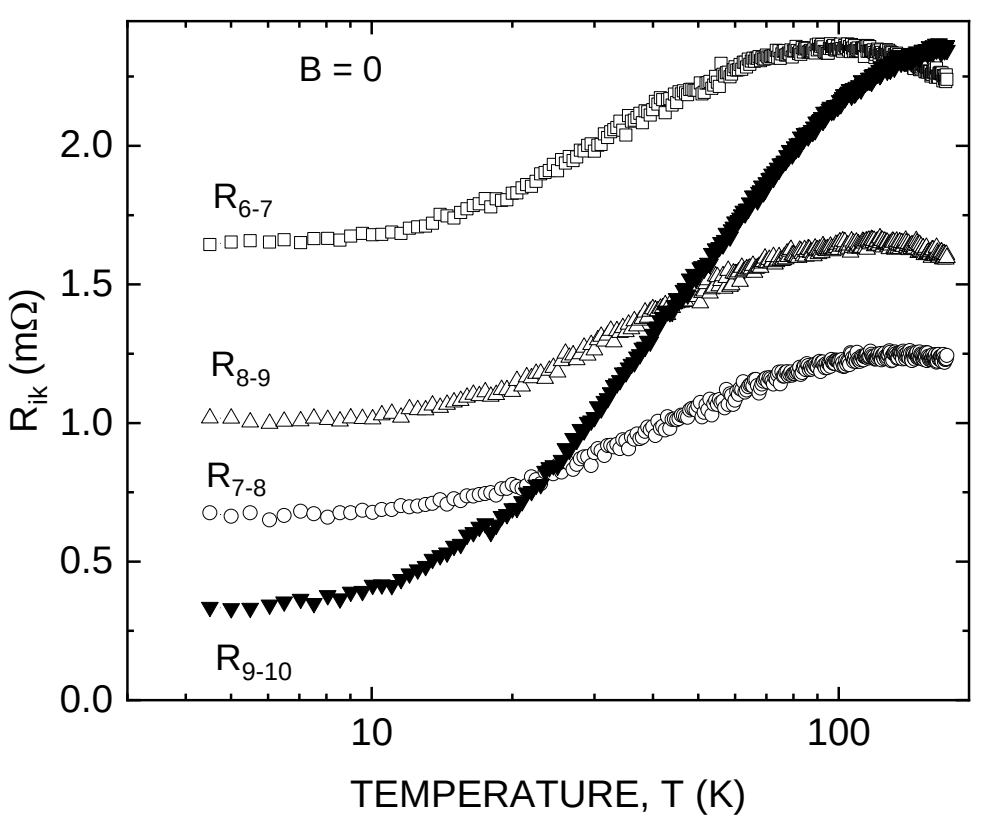

Figure 6. Temperature dependencies of the normal state resistance, $R_{ik}(T) = V_{ik}(T)/I_{1\text{-}11}$, obtained from the measurements in the local configuration, with current flowing between electrodes 1 and 11, $I_{1\text{-}11}$ = 5 mA, and voltages $V_{ik}$(T) measured in the current-applied region.

Because of the structural disorder, the resistance of the LD in the normal state is insulating- or bad metal-like with $dR/dT < 0$.[34] The observation that the relation between the critical current $I_c(T, B)$ and the normal state resistance $R_N(T, B)$ resembles that of the JJ suggests that the LDs can be viewed as chains of the superconducting islands or granules hosting the local superconducting order at $T > 300$ K. The fact that the normal resistance that defines the critical current is the resistance of the bulk graphite, also indicates the crucial role of the non-superconducting bulk graphite substrate that controls establishing the global phase coherence between the SC islands by suppressing the phase slips. This picture is in line with the old proposal by Emery and Kivelson[35] and its development in[36] demonstrates that the metallic layer weakly coupled to the 'pairing layer' with the absent phase stiffness, stabilizes superconductivity and can drive the superconducting transition temperature up to $T_c \approx \Delta_0/2$, where $\Delta_0$ is the preexisting value of the gap in the pairing layer. In our case, this $\Delta_0$ is the zero-temperature gap corresponding to the local intra-granule superconductivity. Taking $T_c \approx 500$ K[8] obtained from superconducting magnetization $M(H)$ hysteresis loops and using for an estimate the BCS result $2\Delta_0/k_B T_c = 3.52$, one arrives at $\Delta_0 \simeq 80$ meV, see[37] for the most recent report on local superconductivity in graphite with $T_c \geq 500$ K. The scanning tunneling spectroscopy measurements performed at the graphite surface at $T = 4.2$ K revealed a superconducting-like gap in the electronic spectrum $\simeq$ 50–80 meV, occurring only in structurally disordered surface regions.[38,39] There are experiments where the dissipative coupling between a conducting layer and the JJ array[40] or 2D films[41] triggers the finite temperature superconductivity. In both cases, the capacitive coupling, which cannot be excluded in our case as well, is behind the phase fluctuations damping. To reveal whether we have capacitive or electron tunneling dissipative coupling, further experiments are required.

### 3.2. Magnetization measurements

Because of the coexistence of localized islands with the ferromagnetic (FM) and SC orders in graphite together with the strong basal-plane diamagnetism, the detection of the small superconducting volume fraction signal by means of the magnetization $M(H, T)$ measurements, which is a standard tool for establishing the existence of superconductivity in conventional bulk superconductors, is the challenging experimental task. Such a task is challenging even in the case of bulk superconducting ferromagnets.[42] Here we report unambiguous evidence for the (i) superconducting magnetization hysteresis loops in the sample with the suppressed ferromagnetism and (ii) magnetization temperature oscillations, as theoretically predicted for SC-FM-SC Josephson junctions. The hysteresis loops measurements in $M(H)$ were performed on the HOPG sample imposed to the heat treatment: first at $T = 800$ K in helium exchange gas, and then in a vacuum about $10^{-2}$ Torr.[8] Such a heat treatment allows for a temporary (because of the aging effect) suppression of the FM response. To reduce the ferromagnetic signal, the sample heat treatment in He, Ar, $N_2$, or just in the low, about 0.05 mbar, vacuum is needed. Together with our observation of the ferromagnetic signal induced or/and enhanced by the graphite oxidation[43], one concludes that the heat treatment leads to the sample deoxidation and hence vanishing or reducing the ferromagnetic signal.

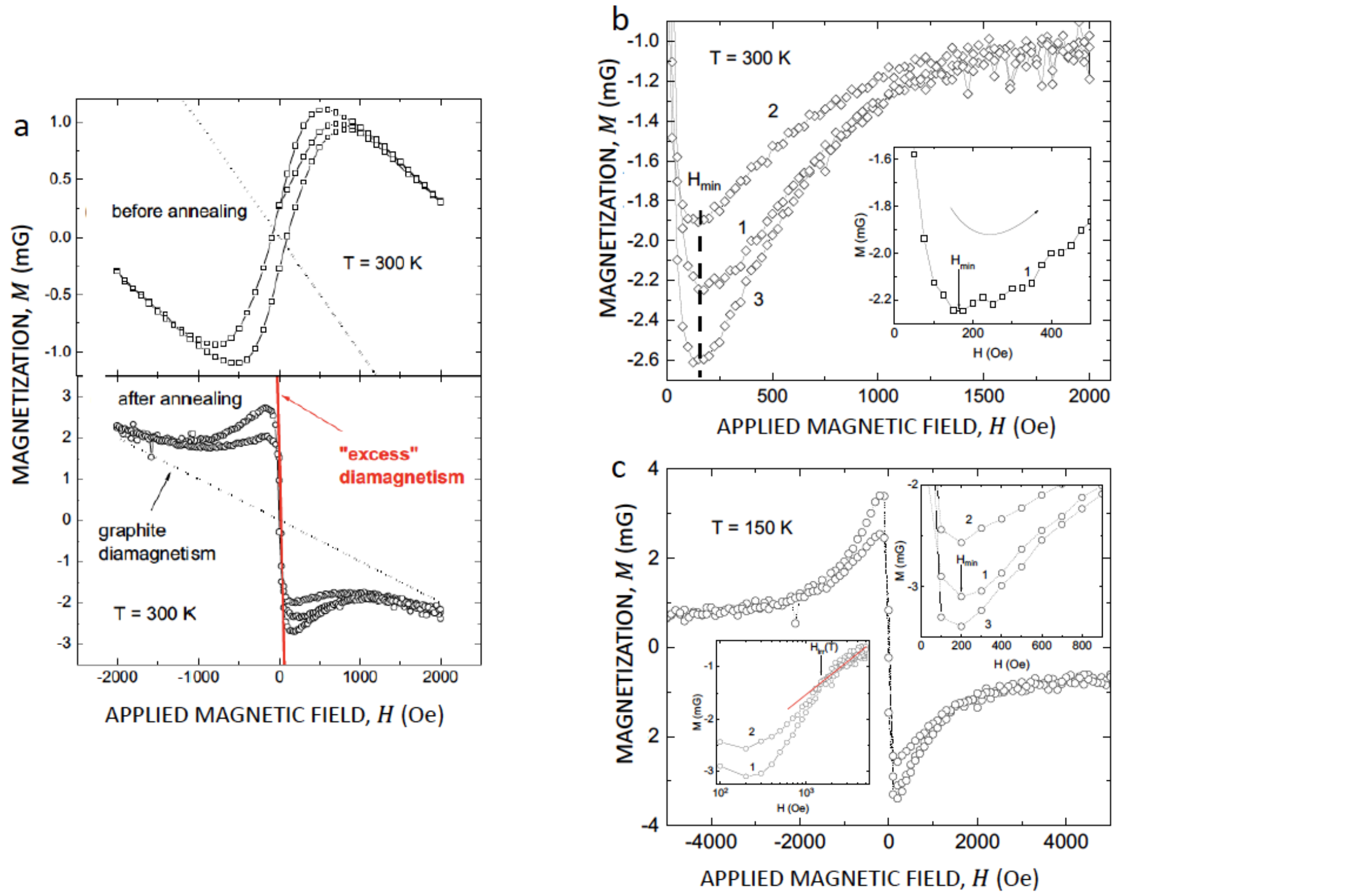

**Figure 7. Magnetization *M*(*H*) hysteresis loops. a:** Magnetization $M(H)$ hysteresis loops measured for HOPG sample of the volume 3.1 x 1.75 x 1.75 mm$^3$ with $H \perp$ c-axis. *Upper panel*: Magnetization before the annealing at $T$=800 K. *Lower panel*: Magnetization taken after the annealing at $T$=800 K. Annealing induces the transformation of the ferromagnetic- to the superconducting-type magnetization behavior. Dotted lines correspond to $M = \chi_{d1}H$ with diamagnetic susceptibility $\chi_{d1} = -10^{-3}$ mG Oe$^{-1}$ for graphite in the $H \perp$ c-axis geometry. Solid line in the lower panel demonstrates the low-field excess diamagnetism with $\chi_{d2} = -8{\cdot}10^{-2}$ mG Oe$^{-1}$. **b:** Magnetization hysteresis loop $M$ ($H > 0$) measured for HOPG at $T$ = 300 K in the field interval – 2 kOe $\leq H \leq$ 2 kOe ($H \perp$ c-axis) after subtraction of diamagnetic signal $M = \chi_{d1}{\cdot}H$, $\chi_{d1} = -10^{-3}$ mG Oe$^{-1}$. Curves 1 and 2 correspond to the ascending (virgin) and descending branches of the $M(H)$, respectively, and curve 3 completes the hysteretic cycle. Inset: the virgin branch of $M(H)$; $H_{min}$ = 170 ± 15 Oe. **c:** The $M(H)$ hysteresis loop obtained for HOPG sample at T = 150 K in the field interval: – 5 kOe $\leq H \leq$ 5 kOe (with $H \perp$ c-axis) after subtraction of the diamagnetic signal $M = \chi_d{\cdot}H$, $\chi_d = -10^{-3}$ mG Oe$^{-1}$. The measurements were performed after cooling the sample in zero applied magnetic field (ZFC regime) towards the target temperature. Upper inset: low-field ($H > 0$) portion of $M$(H); curves 1, 2, and 3 are as defined in panel **b**; arrow indicates $H_{min}$= 220 ± 15 Oe. Lower inset: curves 1 and 2 are as in the upper inset; arrow marks the irreversibility field $H_{irr}$ = 1500 ± 100 Oe; the solid line is the guide for the eye.

The measurements were performed at $T = 300$ K and $T = 150$ K in a magnetic field perpendicular to the graphite crystallographic c-axis, $H \perp$c, at $T = 300$ K and $T = 150$ K. Figure 7a shows the magnetization $M(H)$ measured at $T = 300$ K before, see the upper panel in Fig. 7a, and immediately after, see the lower panel in Fig. 7a, the sample annealing. As clearly shown in Figure 7a, the sample heat treatment transforms the initially FM-type $M(H)$ hysteresis loop into the SC-type one. The dotted lines in Figure 5a are obtained from equation $M = \chi_{d1}H$ with the diamagnetic susceptibility $\chi_{d1} = -10^{-3}$ mG Oe$^{-1}$, characteristic of graphite in the $H \perp$ c-axis geometry[44]. The solid line in the lower panel of Fig. 7a corresponds to equation $M = \chi_{d2}H$ with the diamagnetic susceptibility $\chi_{d2} = -0.08$ mG·Oe$^{-1}$, demonstrating almost two orders of magnitude (by a factor of 80) stronger than the low-field, $H < 100$ Oe, diamagnetism that cannot be accounted for by the diamagnetism of the normal, i.e., non-superconducting graphite. Assuming extra diamagnetism to originate from the Meissner effect, more precisely from the superconducting shielding, one gets the superconducting volume fraction, SF $\approx 0.1$ % of what is expected for an ideal bulk superconductor, being consistent with our conclusion on the superconductivity nucleated within the line defects (LD) at the sample surface.

Figures 7b,c present superconducting-type $M(H)$ hysteresis loops obtained at $T = 300$ K, see Fig. 7b, and $T = 150$ K, see Fig. 7c, after subtraction of the background diamagnetism of the bulk graphite. It is instructive to verify whether the bulk vortex pinning is behind the $M(H)$ hysteresis, as in the Bean model. In this model, $H_{min}(T)$, shown in Figures 7b,c, would be related to the full magnetic field penetration $H_f(T) > H_{c1}(T)$, so that $H_{min} \approx H_f = (1/2)\cdot\mu_0 r j_c$, where $j_c$ is the critical current density, and $r$ is the characteristic sample size or the size of superconducting grains. Taking the height and width of the line defects (LD) as $r$ = 1-10 nm, as a characteristic scale, one estimates $j_c$(T) = $10^8$–$10^9$ A/cm$^2$, being comparable to the characteristic values for depairing critical current density $j_{dep}$=$10^7$–$10^9$ A/cm$^2$. Because our $M(H)$ hysteresis loops do not demonstrate features characteristic for the strong vortex pinning,[23] the observed high values of $j_c$ and hence the realization of the Bean scenario is unlikely. Yet, as Figure 8 shows, $M(H)$ hysteresis loops obtained for HOPG are very similar to the hysteresis loops measured in the $Bi_2Sr_2CaCu_2O_8$ (Bi2212) single crystals, strong type-II superconductors with the superconducting transition temperature $T_c$ = 83 K, at high enough temperatures where the geometrical barrier is the relevant source of $M(H)$ hysteresis in Bi2212.[45]

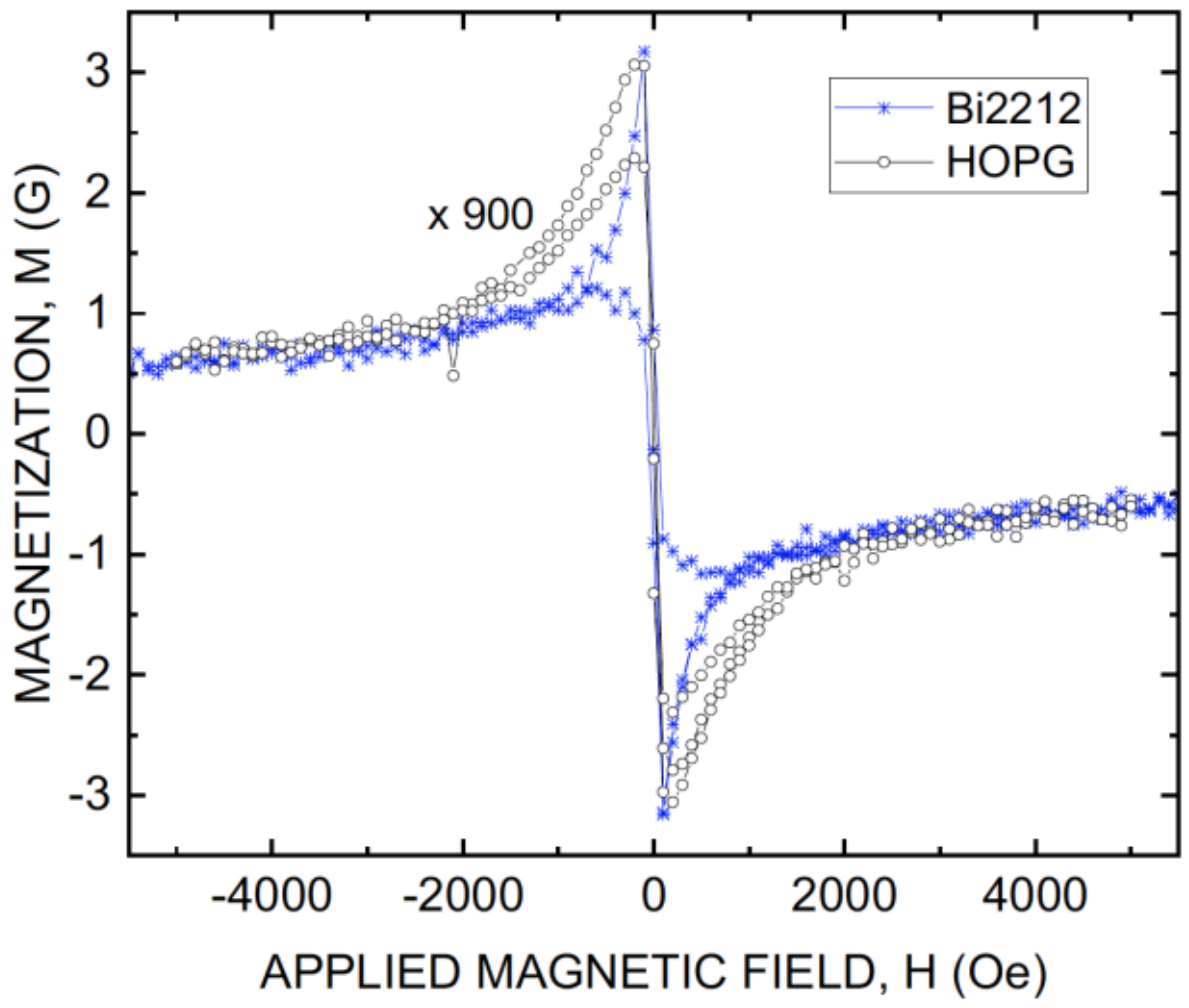

**Figure 8. Magnetization curve.** Magnetization $M(H)$ for superconducting ($T_c$=83 K) $Bi_2Sr_2CaCu_2O_8$ (Bi:2212) single crystal (V = 350 x 350 x 70 μm$^3$), measured with $H$ || c-axis at $T$ = 45 K, and $M(H)$ obtained for HOPG at $T$ = 150 K after subtraction of the diamagnetic background signal and multiplied by factor 900.

Note that $M(H)$ obtained for Bi:2212 and HOPG practically coincide when $M_{HOPG}(H)$ is multiplied by the factor of 900. Taking the superconducting volume fraction SF = 100 % for Bi2212, the factor 900 implies that the SF in the HOPG is about 0.1% which is equal to the SF obtained from the low-field excess diamagnetism data shown in the lower panel of Figure 7a.

Figure 9 presents temperature dependencies of the magnetic moment $m(T)$ measured for the cleaved HOPG sample before the heat treatment, i.e., when superconducting and ferromagnetic contributions coexist. The results of Fig. 9 exemplify the oscillating character of $M(T)$ measured in both ZFC and FCC regimes in several samples. The inset in Fig. 9a shows the measured difference $dm = m_{FCC} - m_{ZFC}$ and also the smoothed line. The observed magnetic moment oscillations well resemble the predicted oscillatory behavior of the Josephson critical current $I_{cJ}(T)$ in SC-FM-SC Josephson junctions[46,47] and perfectly agrees with the “anomalous” temperature dependence of the critical $I_c(T, B)$ obtained from our electrical transport measurements presented in Fig. 2.

Figure 9b illustrates the similarity between the measured oscillations in $dm(T)$ (the upper red curve) and the one theoretically predicted in[47] dependence of $I_{cJ}(T)$ shown in the inset. We stress that Fig. 9b provides convincing experimental evidence for the localized superconductivity in our samples that persists for temperatures as high as $T$ = 300 K, at least.

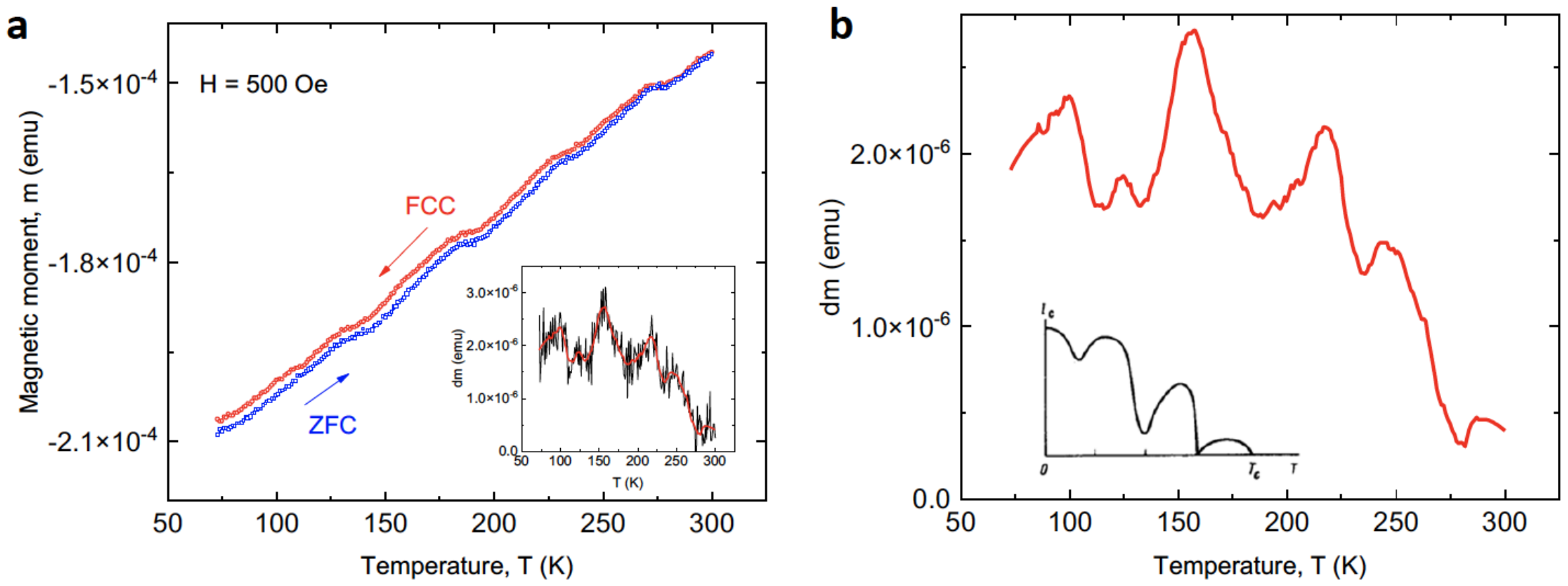

**Figure 9. Magnetic moments. a:** Magnetic moment $m(T, H)$ measured for HOPG with coexisting ferromagnetic and superconducting order parameters in zero-field-cooled (ZFC) and field-cooled on cooling (FCC) regimes in applied magnetic field $H$ (|| c-axis) = 500 Oe. The inset shows measured $dm(T) = m_{FCC}(T) - m_{ZFC}(T)$ (black thin line) and smoothed (thick red line) data. **b:** Smoothed $dm$(T) = $m_{FCC}(T) - m_{ZFC}(T)$ data from the inset in (a). The inset shows the theoretically predicted Josephson critical current $I_{cj}(T)$ oscillations for S-F-S Josephson contacts.[43]

### 3.3. Theory

We now demonstrate, using a simple model, how superconductivity does arise in line defects. Most of the standard existing and proposed superconductivity mechanisms rely on an intermediate boson field mediating an attractive interaction between electrons so that bosonic Cooper pairs form. In standard superconductivity, this boson field originates from lattice vibrations, phonons, see[23] for a comprehensive review. Oscillations of a regular lattice, however, are not the only possible deformations of a solid. In solids containing extended defects, additional oscillation modes localized at these defects[48] also interact with electrons.

The effect of the structural defects in materials on electrons can be described via effective gauge fields.[49] This holds for both strain defects,[50,51] see[52] for a review, and curvature defects.[53,54] In dielectrics, strain gradients cause an electric polarization and vice versa, this is the well-known phenomenon of flexoelectricity.[55] We now show that strain gradient fluctuations can lead to superconductivity in conductors with many free electrons available.

We model a linear defect at the surface of the polycrystalline graphite by a gradient of shear strain across the defect. The effective gauge fields representing strain in single graphite sheets are given by (in this section we use natural units $c = 1$, $\hbar = 1$, $\varepsilon_0 = 1$)

$$A_0 = g\left(u_{xx} + u_{yy}\right)$$
$$A_x = \frac{b}{\ell}\left(u_{xx} - u_{yy}\right)$$
$$A_y = -2\frac{b}{\ell}u_{xy}, \tag{1}$$

where $\ell$ is the lattice constant, $b=\mathcal{O}(1)$ is a numerical factor, $g \approx 4$ eV, and $u$ is the strain tensor.[50-52] Note that, here too, only strain gradients represent effective gauge invariant combinations coupling to electrons. We will consider layers of graphite with the same pure shear strain $u_{xy} = u(x)$, which is a function of $x$ only, corresponding to the unique non-vanishing gauge component $A_y = A(x) = -2(b/\ell)u(x)$. As a specific model we choose the function $u(x)$ as

$$u(x) = \frac{U}{4b}, \qquad x < -w/2,$$
$$u(x) = -U/2bwx, \qquad -w/2 \leq x \leq w/2,$$
$$u(x) = -\frac{U}{4b}, \qquad x > w/2. \tag{2}$$

Here, $U$ is the constant characterizing the shear strain, the line defect (wrinkle) goes along the $y$-axis, and the width of the defect (in the $x$-direction) is $w$. Accordingly, the magnetic field felt by electrons is along the $z$-direction:

$$B = \partial_x A_y(x) = \partial_x A(x) = \frac{U}{\ell w}. \tag{3}$$

The Pauli Hamiltonian for a particle of charge $qe$ and spin $\boldsymbol{s}$ on a 2D sheet with the grain boundary defect is

$$H = \frac{1}{2m}(\boldsymbol{p} - qe\boldsymbol{A}(\boldsymbol{x}))^2 - \frac{qe}{m}\boldsymbol{s} \cdot \boldsymbol{B}(\boldsymbol{x}), \tag{4}$$

where, for the moment, we have neglected the 3D Coulomb interaction $V_C(r) = (qe)^2/(4\pi\varepsilon r)$ with $r$ being the distance between two electrons and $\varepsilon$ being the relative dielectric permittivity of the material. The corresponding time-independent Pauli equation is given by

$$\left[\frac{1}{2m}(\boldsymbol{p}-qe\boldsymbol{A}(\boldsymbol{x}))^2-\frac{qes}{m}B(x)\right]\Psi=E\Psi\,, \tag{5}$$

where $\boldsymbol{A}$ and $B$ are given by Equations (1) and (3), respectively, and $s$ is the $z$-axis spin component.

Since the only non-vanishing component of the gauge potential $\boldsymbol{A}$ is $A_y$, which is a function of $x$ alone, inside the line defect, we can make the Ansatz $\Psi(x,y)=\exp(\mathrm{i}2\pi ky/D)\psi(x)$ with the periodic boundary conditions in the $y$ direction, where $D$ is the sample dimension in the $y$ direction and $k\in\mathbb{Z}$ . Then Eq. (5) reduces to the standard Landau level problem in the Landau gauge

$$\left[-\frac{1}{2m}\partial_x^2+\frac{m}{2}\omega^2\left(x-\frac{2\pi k}{qeBD}\right)^2\right]\psi(x)=(E+s\omega)\psi(x) \tag{6}$$

with the cyclotron frequency $\omega=(qe/m)B$ and $B$ being given by Equation (3). The ground state has $k$=0 and energy $E=((1/2)-s)\omega=0$ and is realized for a configuration with the spin polarized in the direction of the effective magnetic field.

Outside the line defect, where $A_y$ is a constant and $B(x)$=0, the equation for the ground state reduces to

$$\left[\partial_x^2-\left(\frac{qeU}{4\ell}\right)^2\right]\psi(x)=0\,, \tag{7}$$

with the solution

$$\psi\propto\exp(-\lambda|x|),\qquad \lambda=\frac{qeU}{2\ell}\,, \tag{8}$$

showing that the line defect localizes charges within its width. Of course, in a realistic material there are distortions of the effective magnetic field also along the line defects. These can arise, e.g., from modulations of diagonal strain so that $u_{yy}=-u_{xx}$, $A_x(y)=-2(b/\ell)u_{xx}(y)=(Ud/\pi\ell w)\sin(\pi y/d)$. Consequently, the total magnetic field becomes

$$B=\partial_x A_y(x)-\partial_y A_x(y)=\frac{U}{\ell w}\left[1-\cos\left(\frac{\pi}{d}y\right)\right], \tag{9}$$

where $d$=$\mathcal{O}(w)$ is the wavelength of the modulation. Essentially, along the line defect there are now alternating structures of the typical size $d$ of effective magnetic field with the amplitude $B = 2U/\ell w$. If the strain is sufficiently high, so that $w/\ell \gg 1/(2qeU)$, the effective magnetic length satisfies the condition

$$\ell_{\rm mag} = \sqrt{\frac{1}{qeB}} = \sqrt{\frac{\ell w}{qeU}} \ll L \,, \tag{10}$$

and the ground state consists of spin-polarized charges localized by the envelope exponent $\exp(-r^2/4\ell_{\rm mag})$ of the first Landau levels in the regions of high effective magnetic field. Of course, we have used an idealized model as an example; however, the main features of this model do not depend on the exact form of the strain gradients. As soon as inhomogeneous regions of sufficiently strong effective magnetic field form, droplets of localized electrons along the line defects form as well.

Until now, we have considered only the effect of the background strain gradients on electrons. Around these defects, however, small fluctuations of strain gradients also coupling to electrons do exist. Moreover, there is also an unavoidable back reaction effect, the strain fluctuations caused by electrons themselves. Through this coupled effect, electrons will interact with each other by exchanging dynamical strain gradient fluctuations within the droplets of the typical size $L$. These strain gradient fluctuations are modeled as dynamical effective gauge fields minimally coupled to the electrons.

The background network of defects breaks the parity invariance on the graphite sheet and, as a result, the action for the effective gauge field $a_\mu$ (we use lower-case letters for the dynamical, fluctuating component of strain gradient) contains the topological Chern-Simons term[56] as well as the Maxwell term

$$S_{\rm gauge} = \int d^3x \frac{\kappa}{4\pi} a_\mu \epsilon^{\mu\alpha\nu} \partial_\alpha a_\nu - \frac{1}{4g^2} f_{\mu\nu} f^{\mu\nu} \,, \tag{11}$$

where Greek letters denote the components of three-dimensional Minkowski space with coordinates $x$ and time component $x^0 = vt$, with $v = \mathcal{O}(10^{-2})c$ being the velocity of light in graphene, and Einstein notation is used. The quantity $f_{\mu\nu} = \partial_\mu a_\nu - \partial_\nu a_\mu$ is the effective field strength tensor, $\kappa$ is a dimensionless effective coupling, and $g^2$ is an effective coupling with the canonical mass dimension.

Due to the Chern-Simons term, an electron at rest is the source of not only an effective electric field but also of an effective magnetic field. The Pauli interaction of this field with the spin of the second electron results in an attractive component of the pair potential. The generic two-electron problem interacting with gauge fields including the Chern-Simons term has been studied in the late 80s.[57] Upon adding the Coulomb potential $V_c(r) = (qe)^2/(4\pi\varepsilon r)$, the total resulting pair potential for two electrons of mass $m$ and aligned spins, leading to electron pairing and formation of a superconducting ground state, acquires the form

$$V_{\mathrm{tot}} = \frac{\left(L - \frac{1}{\kappa} + \frac{\mu v r}{\kappa} K_1(\mu v r)\right)^2}{m r^2} - \frac{\mu v^{\,2}}{\kappa} \frac{\mu - m}{m} K_0(\mu v r) + V_{\mathrm{c}}(r)\,, \qquad (12)$$

where $L$ is the angular momentum, $\mu = |\kappa| g^2/2\pi$ is the Chern-Simons gauge field mass[56], and $K_{0,1}$ denote modified Bessel functions of the second kind, with the short-distance, $x \ll 1$, asymptotic behaviors

$$K_0 \asymp \ln x\,, \qquad K_1 \asymp \frac{1}{x}\,, \qquad (13)$$

while both functions are exponentially suppressed as $\asymp \exp(-x)$ at large distances, $x \gg 1$.

At large distances, $r \gg 1/v\mu$, only the Chern-Simons term survives. The pair potential then reduces to the Coulomb interaction plus a centrifugal barrier with the effective angular momentum $L - 1/\kappa$. The spectrum remains unchanged for $\kappa = 1/integer$ and, accordingly, this centrifugal barrier is suppressed for $L = 1/\kappa$. This reproduces the well-known result that a spinless charge does not interact with an infinitely thin solenoid. When the Pauli interaction of the magnetic moment is taken into account, however, the resulting singularity must be resolved either by a self-adjoint extension to the Hamiltonian[58] or by considering explicitly the short-distance physics within the solenoid. In both cases, a bound state is shown to exist when the real or the effective, Chern-Simons-induced magnetic moment of the electron is anomalous.[59,60]

In the present case, we have a physical distance scale $w$ set by the line defects width. Strain gradient fluctuations, represented by effective gauge fields, extend to this typical distance, i.e., the gauge field mass is $\mu = \mathcal{O}(1/vw)$. We are thus interested only in the short distance, $r \ll 1/v\mu$, physics below this scale, not in the long-distance behavior. Using the asymptotic behaviors of Eq. (13), one sees right off that, in this case, there is always a minimum in the pair potential at distances $r < 1/v\mu$, if $\mu > m$. Then, the effective magnetic attraction dominates over both effective and the physical Coulomb interactions. The real Coulomb potential becomes the dominant repulsive interaction only at distances $r \gg 1/v\mu$. Since electrons are localized by the background grain boundaries in the above discussed droplets aligned along line defects, the formation of spin-triplet pairs with the consequent local Bose condensation in these regions occurs if the strain gradient fluctuations are sufficiently strong to satisfy the $\mu > m$ condition. The scale of the bound state energy is given by the Chern-Simons mass $\mu = \mathcal{O}(1/wv)$. Taking the typical line defect width $w \approx 10$ nm and the light velocity in graphite $v = 10^6$ m/s, we arrive at the energy scale corresponding to the transition temperature $T_c = \mathcal{O}(1000°)$ K.

If strain gradient fluctuations are sufficiently strong, the system thus forms local droplets of condensate along the defect lines. Global spin-triplet superconductivity can then be established by tunnelling from one droplet to the neighboring one. Furthermore, there may be also tunnelling across the line defects, from one droplet on one line defect to a corresponding droplet on a neighboring line defect. This implies that an irregular, Josephson-junction-array(JJA)-like structure is formed on the surface. The resistance plot, Figure 4a, reproducing the same behaviour as in thin superconducting films, suggests that this is indeed the case and that this JJA surface structure is in its Bose metal state,[61,62] with the charge transport percolation edges forming exactly along the line defects and where the applied magnetic field plays the role of the parameter driving the quantum phase structure, see[63] for a review.

The metallic resistance saturation at low temperatures is caused by quantum phase slip instantons[64] in the effective Josephson junction chains (JJC) along the edge defects. In the limit in which the charging energy is dominated by the ground capacitances, JJC are described by the compact version of the global O(2) model[63] for angles $\varphi$

$$S = g \sum_x d_\mu \varphi d_\mu \varphi \,, \tag{14}$$

where $g^2 = E_J/2E_C$, $E_J$ and $E_C$ are the Josephson coupling and the charging energy, respectively, and $x$ denotes the points of a 2D Euclidean lattice. Quantum phase slips are the instantons of this model,[64] represented by integers $m_x$, and described by a Euclidean 2D Coulomb gas action

$$S_{\mathrm{QPS}} = 4\pi^2 g \sum_x m_x \frac{1}{-\nabla^2} m_x \,. \tag{15}$$

When $g$ is sufficiently large, thus, the instantons undergo a Berezinskii-Kosterlitz-Thouless transition, and quantum phase slips are suppressed.[65] In the present case, the role of $g$ is played by the inverse magnetic field. When this inverse field becomes sufficiently small, the system becomes superconducting along the line defects. Our results thus establish an important result that at zero temperature, $T = 0$, superconductivity is established in 1D systems in accordance with considerations presented in the review.[64]

At finite temperatures, the 1D superconductivity is disrupted by thermal phase slips even at low temperatures.[64] Notably, however, these line defects, arrays of wrinkles and steps at the surface of the cleaved graphite are not independent quantum wires. As they are embedded in the surface of a bulk material, the phase slips disrupting superconductivity hosted by the defects are nothing but the surface point vortices crossing the defect. However, these point vortices continue into the metallic underlying graphite until they spread over the metallic volume and, therefore, are to be viewed as the endpoints of 1D vortex lines extended into the bulk. Since the mobility of these bulk vortices is severely hampered by small resistance $R_N$, the motion of point vortices on the surface is also impeded, and, correspondingly, the phase slips across the line defects are strongly suppressed in this dimensional cascade. This implies an effective increase of $g$ in the JJC on the line defects, which leads to a notable increase in their critical temperature. Therefore, the 1D superconductivity that we report here is the superconductivity of not the independent quantum wires but that of line defects forming at the surface. This implies the high critical temperature $T_c$ arising from the stabilizing effect due to their interaction with the bulk material where they live. This stabilizing effect of the bulk Bernal graphite guarantees that the superconducting state survives at room temperature.

Let us conclude this section by pointing out that, as already noted,[57] the pairing symmetry is the $p$-wave if the value $\kappa = 1$, so that the centrifugal barrier is completely canceled at large distances for $L = 1$. In this case, half-quantum vortices have been shown to exist,[66-68] and their contribution results in the appearance of the observed metastable state causing the resistance plateau at $R_N/2$ shown in Figure 4.

## 4. Conclusion

We have reported the first-ever observation of the global room-temperature superconductivity at ambient pressure. Notably, while a single graphite layer, graphene, is hailed as a miracle material of the new century, the bulk pyrolytic graphite opens the way to even more spectacular advances in technology. The experimental data clearly demonstrated that the array of nearly parallel linear defects that form due to the cleaving of the highly oriented pyrolytic graphite hosts one-dimensional superconductivity. Our measurements at the ambient pressure at temperatures up to 300 K and applied magnetic field perpendicular to the basal graphitic planes up to 9 T, revealed that the superconducting critical current $I_c(T, B)$ is proportional to $1/R_N(T, B)$, indicating the Josephson-junction like nature of the emerging superconductivity. This latter conclusion is supported by the oscillations of the critical current with temperature that are characteristic of superconductor-ferromagnet-superconductor Josephson junctions. Global superconductivity arises due to global phase coherence in the superconducting granules array promoted by the stabilizing effect of underlying Bernal graphite having the resistance $R_N$. Our theory of global superconductivity emerging on the array of linear structural defects well describes the experimental findings.

The ideas and concepts explored in our work are not confined to graphite. Our theoretical model is quite general and guides where to look for more room-temperature superconducting materials. The basic principle we have uncovered is that linear defects in stacked materials host strong strain gradient fluctuations, which induce the local pairing of electrons into condensate droplets that form JJA-like structures in the planes. The global superconductivity is then established by the effect of the tunneling links connecting the superconducting droplets. If the droplets are sufficiently small, one foresees a fairly high critical superconducting temperature.

**5. Experimental Section.**

The multi-terminal basal-plane resistance $R_b(T, B, I)$ measurements are taken on the freshly cleaved HOPG sample with the dimensions $l \cdot w \cdot t$ = 5·4 ·0.5 mm$^3$, obtained from the Union Carbide Co., using Janis 9 T magnet He-cryostat, Keithley 6220 Precision Current Source, Keithley 2182A Nanovoltmeter, and Lake Shore 340 Temperature Controller. The studied samples possess the out-of-plane/basal-plane resistivity ratio $\rho_c/\rho_b \approx 3 \cdot 10^4$ at $T$ = 300 K and $B$ = 0, with $\rho_b$ = 5 μΩ·cm. The X-ray diffraction (Θ-2Θ) spectra of the virgin HOPG samples demonstrate characteristic hexagonal graphite structure with no signatures of other phases. The crystal lattice parameters are $a$ = 2.48 Å and $c$ = 6.71 Å. The high degree of crystallite orientation along the hexagonal $c$-axis is confirmed by the X-ray rocking curve measurements with the full width at half maximum (FWHM) = 0.3°. The arrays of nearly parallel surface wrinkles – linear defects – were produced by the mechanical exfoliation of graphene surface layers from the bulk HOPG sample using the commercially available Scotch tape, which was stuck with finger pressure to the sample surface and then exerted a force normal to the sample surface. The cleavage procedure was repeated several times to get as large as possible planar areas. The sample surface topography is analyzed by means of Atomic Force Microscopy and Scanning electron microscopy on 10 x 10 $\mu m^2$ and 100 x 100 $\mu m^2$ areas, respectively.

The resistance $R(T, B, I)$ measurements are performed in the 4.5 K $\leq T \leq$ 300 K temperature interval, $0 \leq B \leq 9$ T applied magnetic fields range with $B$||c-axis, and the dc electric current $I \geq 5$ μA. In experiments, we use the line-electrode geometry to measure the in-plane resistance in both local and non-local configurations. The magnetization, $M(T, B)$, measurements were performed for $B$ || c-axis and $B \perp$ c-axis magnetic field configurations in the fields up to 7 T and temperatures between 2 and 300 K by means of the SQUID MPMS7 magnetometer (Quantum Design).

**Acknowledgements**

The work was supported by FAPESP, CNPq, and Terra Quantum AG. YVK and VMV conceived the research, YVK, JHST, RRdS, and FSO performed the measurements, YVK, VMV, MCD, and CAT analyzed and interpreted the data, MCD, CAT, and VMV composed a theory, YVK, VMV, MCD, and CAT wrote the paper, all authors discussed the manuscript, YVK supervised the experimental work. We are delighted to thank G. Baskaran, A. Goldman, V. A. Khodel, and V. P. Mineev for valuable discussions.

**Conflict of Interest**

The authors declare that the research was conducted in the absence of any commercial or financial relationships that could be construed as a potential conflict of interest.

**Data Availability Statement:**

The data that support the findings of this study are available upon request.